\documentclass[12pt]{article}
\usepackage{titlesec}
\usepackage{lineno}
\usepackage{lipsum}
\usepackage{changepage}
\usepackage{graphicx}
\usepackage{amsmath}
\usepackage{caption}
\usepackage{setspace}
\usepackage{color}
\usepackage{comment}
\usepackage{authblk}
\usepackage{bm}

\usepackage[numbers,sort&compress]{natbib}

\usepackage{times}

\usepackage[
 left=3cm,  
 right=2.6cm,  
 top=2.1cm,   
 bottom=2.1cm, 
]{geometry}

\definecolor{qired}{rgb}{0.6, 0, 0}

\DeclareCaptionFormat{boldnumberfirstsentence}{
 \textbf{#1#2}#3\par
}

\DeclareCaptionLabelSeparator{colonspace}{: }
\newcommand*\patchAmsMathEnvironmentForLineno[1]{%
 \expandafter\let\csname old#1\expandafter\endcsname\csname #1\endcsname
 \expandafter\let\csname oldend#1\expandafter\endcsname\csname end#1\endcsname
 \renewenvironment{#1}%
   {\linenomath\csname old#1\endcsname}%
   {\csname oldend#1\endcsname\endlinenomath}}%
\newcommand*\patchBothAmsMathEnvironmentsForLineno[1]{%
 \patchAmsMathEnvironmentForLineno{#1}%
 \patchAmsMathEnvironmentForLineno{#1*}}%
\AtBeginDocument{%
\patchBothAmsMathEnvironmentsForLineno{equation}%
\patchBothAmsMathEnvironmentsForLineno{align}%
\patchBothAmsMathEnvironmentsForLineno{flalign}%
\patchBothAmsMathEnvironmentsForLineno{alignat}%
\patchBothAmsMathEnvironmentsForLineno{gather}%
\patchBothAmsMathEnvironmentsForLineno{multline}%
}

\allowdisplaybreaks

\title{Direct reciprocity in asynchronous interactions}
\date{}

\author[1]{\fontsize{12}{14}\selectfont Ketian Sun}
\author[2,3,4,*]{\fontsize{12}{14}\selectfont Qi Su}
\author[1,5,*]{\fontsize{12}{14}\selectfont Long Wang}
\affil[1]{Center for Systems and Control, College of Engineering, Peking University, Beijing, 100871, China.}
\affil[2]{Department of Automation, Shanghai Jiao Tong University, Shanghai, 200240, China}
\affil[3]{Key Laboratory of System Control and Information Processing, Ministry of Education of China, Shanghai, 200240, China}
\affil[4]{Shanghai Engineering Research Center of Intelligent Control and Management, Shanghai, 200240, China}
\affil[5]{Center for Multi-Agent Research, Institute for Artificial Intelligence, Peking University, Beijing, 100871, China}
\affil[*]{Corresponding author. Email: \text{qisu@sjtu.edu.cn}; \text{longwang@pku.edu.cn}}

\begin{document}
\maketitle

\begin{abstract}
Cooperation is vital for the survival of living systems but is challenging due to the costs borne by altruistic individuals. Direct reciprocity, where actions are based on past encounters, is a key mechanism fostering cooperation. However, most studies assume synchronous decision-making, whereas real-world interactions are often asynchronous, with individuals acting in sequence. This asynchrony can undermine standard cooperative strategies like Tit-for-Tat and Win-Stay Lose-Shift. To better understand cooperation in real-world contexts, it is crucial to explore the theory of direct reciprocity in asynchronous interactions. To address this, we introduce a framework based on asynchronous stochastic games, incorporating asynchronous decisions and dynamic environmental feedback. We analytically derive the conditions under which strategies form cooperative Nash equilibria. Our results demonstrate that the order of interactions can significantly alter outcomes: interaction asynchrony generally inhibits cooperation, except under specific conditions where environmental feedback effectively mitigates its negative impact. When environmental feedback is incorporated, a variety of stable reciprocal strategies can be sustained. Notably, above a critical environmental threshold, any cooperative strategy can form a Nash equilibrium. Overall, our work underscores the importance of interaction order in long-term evolutionary processes and highlights the pivotal role of environmental feedback in stabilizing cooperation in asynchronous interactions.
\end{abstract}

\newpage
\section{Introduction}
Cooperation is essential for the survival of diverse living systems, including microbial communities \cite{allen2013spatial, nadell2016spatial}, animal groups \cite{wilkinson1984reciprocal, voelkl2015matching, silk2013chimpanzees}, and human societies \cite{melis2010human, rand2013human}. However, cooperation involves altruistic behavior that incurs costs to benefit others, potentially compromising one's immediate interests and placing the individual at a disadvantage. Understanding the emergence and maintenance of cooperation has long posed an evolutionary challenge \cite{hardin1968tragedy, efferson2024super, hauser2014cooperating, nowak2010evolution}. Over recent decades, numerous studies have delved into this topic, proposing various mechanisms to elucidate cooperative behavior. These mechanisms include kin selection \cite{taylor1992altruism, rousset2000theoretical}, direct reciprocity \cite{mcavoy2022evolutionary, tkadlec2023mutation}, indirect reciprocity \cite{ghang2015indirect, okada2018solution, ohtsuki2006leading}, network reciprocity \cite{ohtsuki2006simple, allen2017evolutionary, su2023strategy, sheng2024strategy, su2022evolution, su2022evolution2, mcavoy2020social, li2020evolution}, and group selection \cite{cooney2023evolutionary, pfeiffer2005evolution}. Among these mechanisms, direct reciprocity, characterized by repeated interactions where players make decisions based on past encounters, has garnered significant attention in both theoretical and experimental research \cite{efferson2024super,nowak2004emergence, hilbe2017memory, baek2016comparing, laporte2023adaptive, hilbe2014cooperation, chen2023outlearning, hilbe2018partners, hauert1997effects, reiter2018crosstalk, stewart2013extortion, pinheiro2014evolution}.
For instance, within the framework of direct reciprocity, several well-known strategies have been proposed and extensively studied for their effectiveness in sustaining cooperation. These include Tit-for-Tat (TFT) \cite{axelrod1981evolution}, Generous Tit-for-Tat (GTFT) \cite{nowak1992tit}, and Win-Stay Lose-Shift (WSLS) \cite{nowak1993strategy}. Additionally, Press and Dyson introduced the concept of ``zero-determinant'' (ZD) strategies, which enable players to enforce a linear relationship between their own payoffs and those of their co-players \cite{press2012iterated}. Furthermore, Akin analyzed all Nash equilibria that support full cooperation, referred to as partner strategies \cite{akin2015you, hilbe2018partners}.

The existing theory of direct reciprocity has provided valuable insights into the emergence and maintenance of cooperation. However, this theory is primarily based on the assumption of synchronous interactions, where actions occur synchronously. In contrast, many reciprocal actions among species are asynchronous. Notable examples include vampire bats sharing blood \cite{wilkinson1984reciprocal}, primates engaging in social grooming \cite{dunbar1991functional}, and pied babblers taking turns as sentinels \cite{ridley2013sentinel}.
The distinction between synchronous and asynchronous interactions, though seemingly simple, significantly impacts decision-making processes. In economic models like the Cournot duopoly, the firm that acts first gains an advantage by achieving higher profits compared to scenarios where both firms act synchronously. Synchronous interactions assume that both players are treated equally in terms of timing, whereas asynchronous interactions introduce a form of heterogeneity, as one player must act first. Additionally, in synchronous interactions, individuals make decisions without knowledge of the other's actions; in asynchronous interactions, one player makes his decision based on the co-player's prior action. Previous studies on direct reciprocity have demonstrated that rational decisions in synchronous and asynchronous contexts can differ markedly. Strategies such as Tit for Tat and Win-Stay Lose-Shift, which are effective in synchronous games, may falter in asynchronous contexts due to misaligned moves \cite{nowak1994alternating, frean1994prisoner}. This highlights the necessity for more forgiving strategies \cite{zagorsky2013forgiver}.
Given the prevalence of asynchronous interactions and the substantial impact of timing and information on decision-making, exploring the theory of direct reciprocity in asynchronous contexts is crucial \cite{mcavoy2017autocratic, park2022cooperation}.

In addition to considering asynchronous interactions, many previous studies on direct reciprocity have assumed that the environments in which individuals operate remain constant over time. This assumption, suggesting that individuals participate in a single, fixed game, often oversimplifies the complexities of real-world scenarios \cite{tilman2020evolutionary, su2019evolutionary, wang2021evolution}. Numerous experimental studies have shown that environmental conditions frequently change, impacting how individuals interact.
For instance, overgrazing can lead to the degradation of shared pasture lands, leaving herders with diminished resources in subsequent seasons \cite{aderinto2020can}. By keeping livestock numbers within sustainable limits, herders can promote responsible land use. In these situations, the actions of individuals directly affect the state of the environment, which in turn influences the decisions of the community.
Moreover, environmental changes can stem from both internal factors, such as individual behavior, and external factors, like seasonal variations in climate and soil conditions. This phenomenon is not exclusive to human activities; it can also be observed in animal behaviors. For example, beavers significantly alter ecosystems through dam construction \cite{gurnell1998hydrogeomorphological}, while elephants reshape savanna landscapes by uprooting trees \cite{coverdale2016elephants}. Despite the importance of these processes, the outcomes of asynchronous interactions within dynamic environments remain largely unexplored.

In this paper, we introduce a framework for asynchronous stochastic games, where two players choose whether to cooperate or defect. Cooperation entails incurring a cost to provide a benefit determined by the current environmental state, while defection involves avoiding both the cost and the provision of any benefit. The environment itself evolves in response to players' actions, switching between states that offer varying levels of benefits for cooperation. 
We analytically derive the conditions for strategies to be cooperative Nash equilibria in such asynchronous stochastic games. Our findings show that the order of interactions significantly influences cooperation rates: asynchronous decision-making generally hinders the evolution of cooperation compared to synchronous ones. 
Nonetheless, our findings also identify significant exceptions, which underscore the potential of designing  transition rules to promote cooperation despite the challenges posed by interaction asynchrony. Furthermore, environmental feedback can promote the stability of cooperative strategies in asynchronous decision-making processes. Notably, when the benefit difference between environmental states surpasses the cost of cooperation, any cooperative strategy can be a Nash equilibrium, which would never be observed in static environment contexts. Our work highlights the importance of interaction order in long-term evolutionary processes and the critical role of environmental feedback in stabilizing cooperation in asynchronous interactions.

\section{Model}
\begin{figure}[!t]
	\centering
	\includegraphics[width=.8\textwidth]{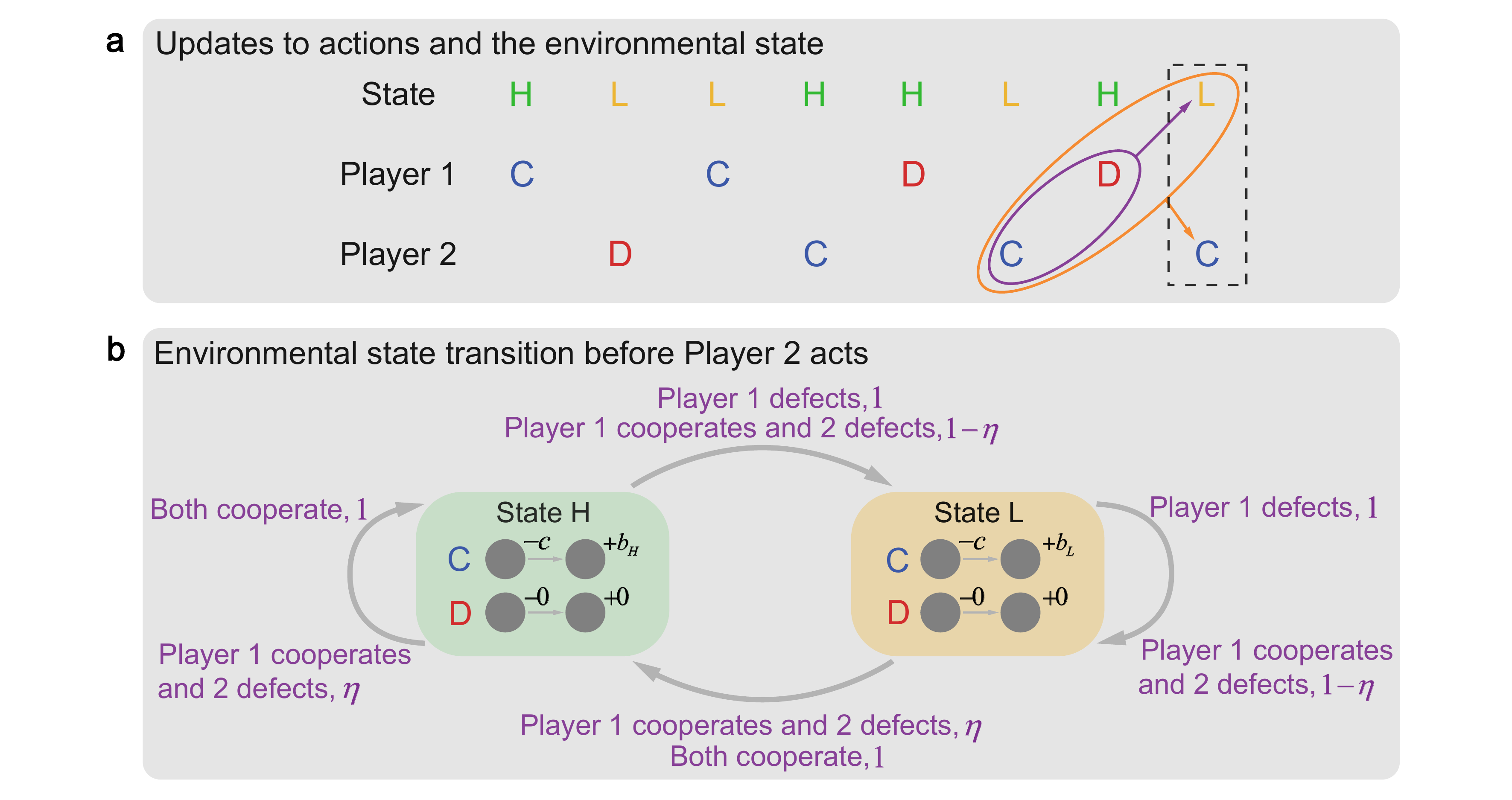}	\caption{\textbf{Asynchronous interactions with environmental feedback.}  \textbf{a}, Updates to each player's actions and the environmental state.
    During each period, the environmental state transitions based on the players' actions from the previous two periods (illustrated by the purple ellipse). For instance, Player 1's defection (D) in the last period combined with Player 2's cooperation two periods ago results in a transition of the environment from a high-benefit state (H) to a low-benefit state (L).
	A player's choice of action is influenced by the current environmental state, his own action from two periods ago, and his co-player's action in the last period (illustrated by the orange ellipse).
     \textbf{b}, An example of the environmental state transition pattern.
	When Player 1 defects in the last period and Player 2 cooperates two periods ago, the environment must transition to a low-benefit state. However, if Player 1 cooperates and Player 2 defects, the environment transitions to a high-benefit state with probability $\eta$; otherwise, it deteriorates.
     }
	\label{fig_1}
\end{figure}

In the main text, we examine a simple asynchronous stochastic game, with further extensions discussed in SI Appendix section 1. The model involves an infinitely repeated game between two players, Player 1 and Player 2, who interact within an environment comprising two states: high (H) and low (L). Each round consists of two periods (see Fig.~\ref{fig_1}\textbf{a}), where Player 1 acts first, followed by Player 2. In each period, a player chooses either cooperation (C) or defection (D) in the current environmental state $s$. Cooperation incurs a cost $c$ and provides a benefit $b_s$ to the co-player, while defection incurs no cost and provides no benefit. We assume $b_H \geq b_L$, indicating that state H has more resources than state L, and define $\Delta = b_H - b_L$ to quantify this resource disparity. The asynchronous static games corresponding to states H and L are referred to as high-benefit and low-benefit games, respectively.

The environmental state transitions over time based on players' actions in the previous two periods, described by a transition vector (see Fig.~\ref{fig_1}\textbf{b}):
\begin{equation}
    \bm{\lambda}=(\lambda_{CC}, \lambda_{CD}, \lambda_{DC}, \lambda_{DD}),
\end{equation}
where $\lambda_{a\tilde{a}} \in [0,1]$ represents the probability that the environment transitions to state H, which depends on the focal player's action $a$ two periods ago and the co-player's action $\tilde{a}$ in the last period. For instance, $\lambda_{CC}$ denotes the probability of transitioning to state H when both players cooperated in their respective preceding turns.
While the framework allows for the exploration of arbitrary dynamic environments, we focus on two real-world scenarios: (1) cooperation improves the quality of environmental resources, while defection causes degradation; (2) the impact of a player's actions diminishes over time, making recent cooperation more influential than earlier actions in promoting high-quality environmental states.
To reflect these dynamics, we impose the ordering $\lambda_{CC} \geq \lambda_{DC} \geq \lambda_{CD} \geq \lambda_{DD}$. For simplicity, our analysis primarily centers on the specific transition vector $\bm{\lambda} = (1, 0, \eta, 0)$.

Each player makes a decision based on the current environmental state, his own action from two periods ago, and his co-player's action in the last round (illustrated by the orange ellipse in Fig.~\ref{fig_1}\textbf{a}). A player's memory-1 strategy can be represented as:
\begin{equation}
	\mathbf{p}:=\left(p_{C C}^{H}, p_{C D}^{H}, p_{D C}^{H}, p_{D D}^{H}, p_{C C}^{L}, p_{C D}^{L}, p_{D C}^{L}, p_{D D}^{L}\right),
\end{equation}
where $p_{a\tilde{a}}^s \in [0, 1]$ is the probability that a player chooses to cooperate, given the environment state $s$, his own action $a$ from two periods ago and his co-player's action $\tilde{a}$ in the last period.
If a strategy depends solely on the environmental state and the co-player's action, and is independent of the player's own previous actions, it is termed reactive.
The strategy $\mathbf{p}$ is reactive if the following conditions are met:
\begin{equation}
	p_{C C}^{H}=p_{D C}^{H},\quad p_{C D}^{H}=p_{D D}^{H},\quad p_{C C}^{L}=p_{D C}^{L},\quad p_{C D}^{L}=p_{D D}^{L}.
\end{equation} 
Additionally, we account for implementation errors, where players occasionally take the opposite action with a probability  $\varepsilon$. In scenarios where the environment remains static, our model reduces to an asynchronous static game \cite{park2022cooperation}. Under such conditions, the following simplifications apply: 
\begin{equation}\label{recover}
	\begin{aligned}
		&b_H=b_L=: b,\\
		&p_{C C}^{H}=p_{C C}^{L}=:p_{C C},\quad p_{C D}^{H}=p_{C D}^{L}=:p_{C D},\\
		&p_{D C}^{H}=p_{D C}^{L}=:p_{D C},\quad p_{D D}^{H}=p_{D D}^{L}=:p_{D D}.
	\end{aligned}
\end{equation}

\section{Results}
In this section, we begin by analyzing partner strategies in asynchronous stochastic games, including deriving expressions for these strategies and examining the prevalence of partner strategies across all cooperative strategies. We then explore how dynamic factors, such as the differences between the two states, influence partner strategies. Next, we consider an evolving population of individuals and investigate the evolutionary outcomes in asynchronous stochastic games. Finally, we discuss the impact of interaction order—asynchronous versus synchronous—on cooperation, highlighting the key differences in evolutionary outcomes across these interaction modes.

\subsection*{Partner strategies in asynchronous stochastic games}

We first investigate which memory-1 strategies qualify as partner strategies in asynchronous stochastic games without errors. Partner strategies achieve full cooperation in self-play and constitute Nash equilibria \cite{hilbe2018partners}. Specifically, when both players adopt a partner strategy $\mathbf{p}$, they establish long-term stable cooperation, ensuring that neither player has an incentive to deviate from this strategy. This sustained cooperation is essential for optimizing outcomes for both players and for the sustainable management of shared resources in asynchronous interactions within dynamic environments.

\begin{figure}[!t]
	\centering
	\includegraphics[width=.8\textwidth]{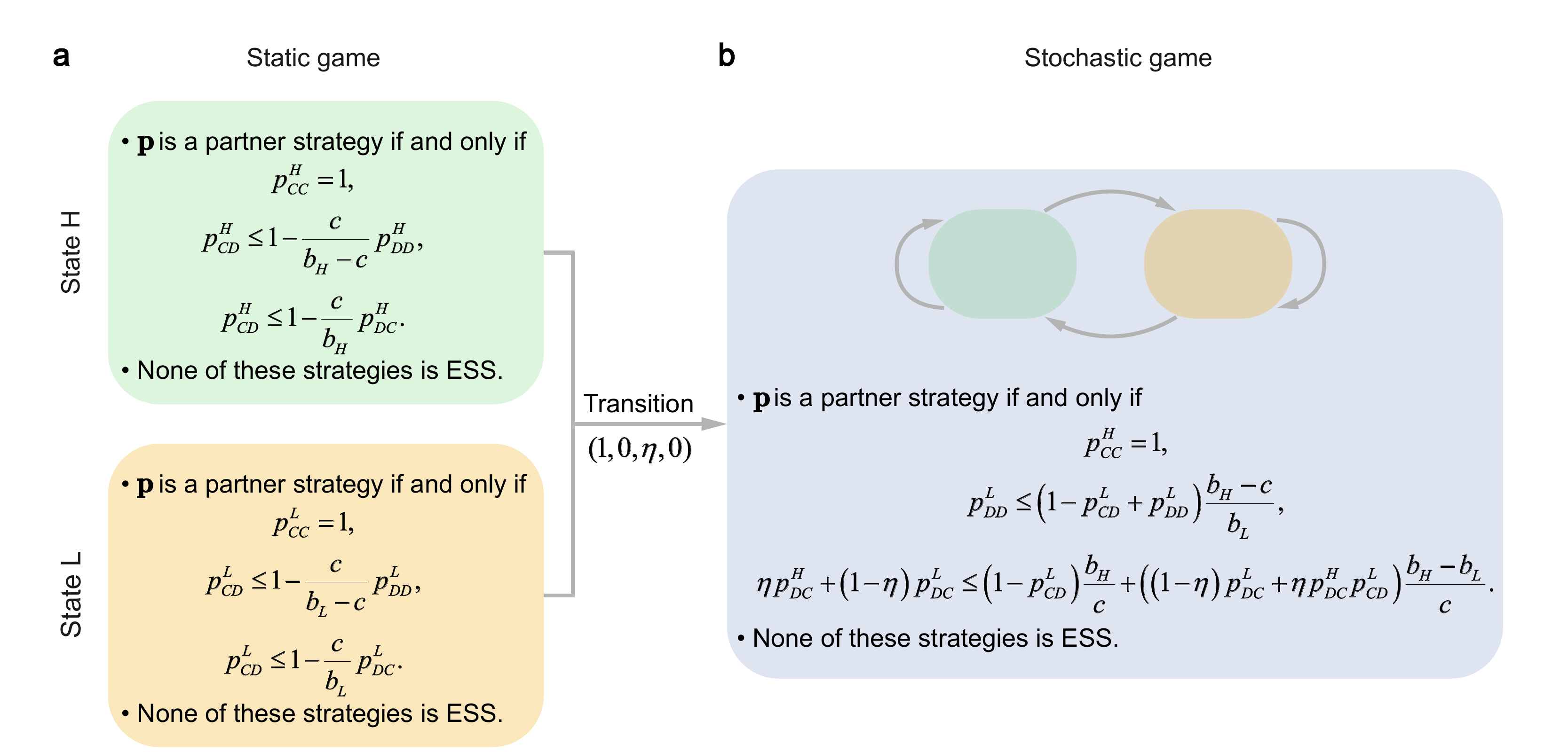}
	\caption{\textbf{Partner strategies in asynchronous static and stochastic games.} 
    \textbf{a}, Partner strategies in asynchronous static high-state games (green region) and asynchronous static low-state games (brown region) with $\varepsilon=0$.
    \textbf{b}, Partner strategies in asynchronous stochastic games, with the state transition vector $\bm{\lambda}=(1,0,\eta,0)$. }
	\label{fig_2}
\end{figure}

Through theoretical analysis (see SI Appendix section 3 for details), we derive the conditions for $\mathbf{p}$ to  be qualified as a partner strategy, as confirmed by the simulations in Fig. S1. The conditions are given by:
\begin{equation}\label{partner_condition}
	\begin{aligned}
		&p_{CC}^{H}=1, \\
		&p_{DD}^{L} \leq\left(1-p_{CD}^{L}+p_{DD}^{L}\right) \frac{b_{H}-c}{b_{L}}, \\
		&\eta p_{DC}^{H}+(1-\eta) p_{DC}^{L} \leq\left(1-p_{CD}^{L}\right) \frac{b_{H}}{c}\\
        &\quad \quad \quad\quad\quad\quad\quad\quad\quad\quad+\left((1-\eta) p_{DC}^{L}+\eta p_{DC}^{H} p_{CD}^{L}\right) \frac{b_{H}-b_{L}}{c}.
	\end{aligned}
\end{equation}
These conditions are independent of $p_{CD}^H$, $p_{DD}^H$, and $p_{CC}^L$, since transitions to the high-benefit state following defection or the low-benefit state after mutual cooperation are not feasible. By adjusting parameters according to Eq.~(\ref{recover}), we can recover the partner strategy criteria for asynchronous static games \cite{park2022cooperation}:
\begin{equation}\label{partner_condition_static}
	\begin{aligned}
            &p_{CC} = 1,\\
		&p_{CD} \le 1-\frac{c}{b-c}p_{DD},\\
		&p_{CD} \le 1-\frac{c}{b}p_{DC}.
	\end{aligned}
\end{equation}

\subsection*{Stability of cooperative strategies in asynchronous stochastic games}
In the previous subsection, we presented the expressions for partner strategies in both asynchronous stochastic and static games. Building on this foundation, we now investigate the abundance of partner strategies in these two types of scenarios. 
Specifically, we focus on the space of cooperative strategies, where mutual cooperation must lead to cooperative actions in the next period in asynchronous stochastic games ($p_{CC}^H = 1$), or in the next round in asynchronous static games ($p_{CC} = 1$).
A higher prevalence of partner strategies, therefore, signifies greater stability for cooperative strategies and more favorable outcomes for the entity involved.
To measure the prevalence of partner strategies within this cooperative space, we adopt the following approach: (1) Fix $p_{CC}^H = 1$ in the asynchronous stochastic games (or $p_{CC} = 1$ in the asynchronous static games); (2) Sample all other elements $p_{a\tilde{a}}^s$ uniformly within the interval [0,1] in the asynchronous stochastic games (or $p_{a\tilde{a}}$ uniformly within [0,1] in the asynchronous static games); (3) Count the number of strategies that satisfy Eq.~(\ref{partner_condition}) (or Eq.~(\ref{partner_condition_static})), and analyze the frequency of such strategies across all sampled cooperative strategies.

\begin{figure}[!t]
	\centering
	\includegraphics[width=.8\textwidth]{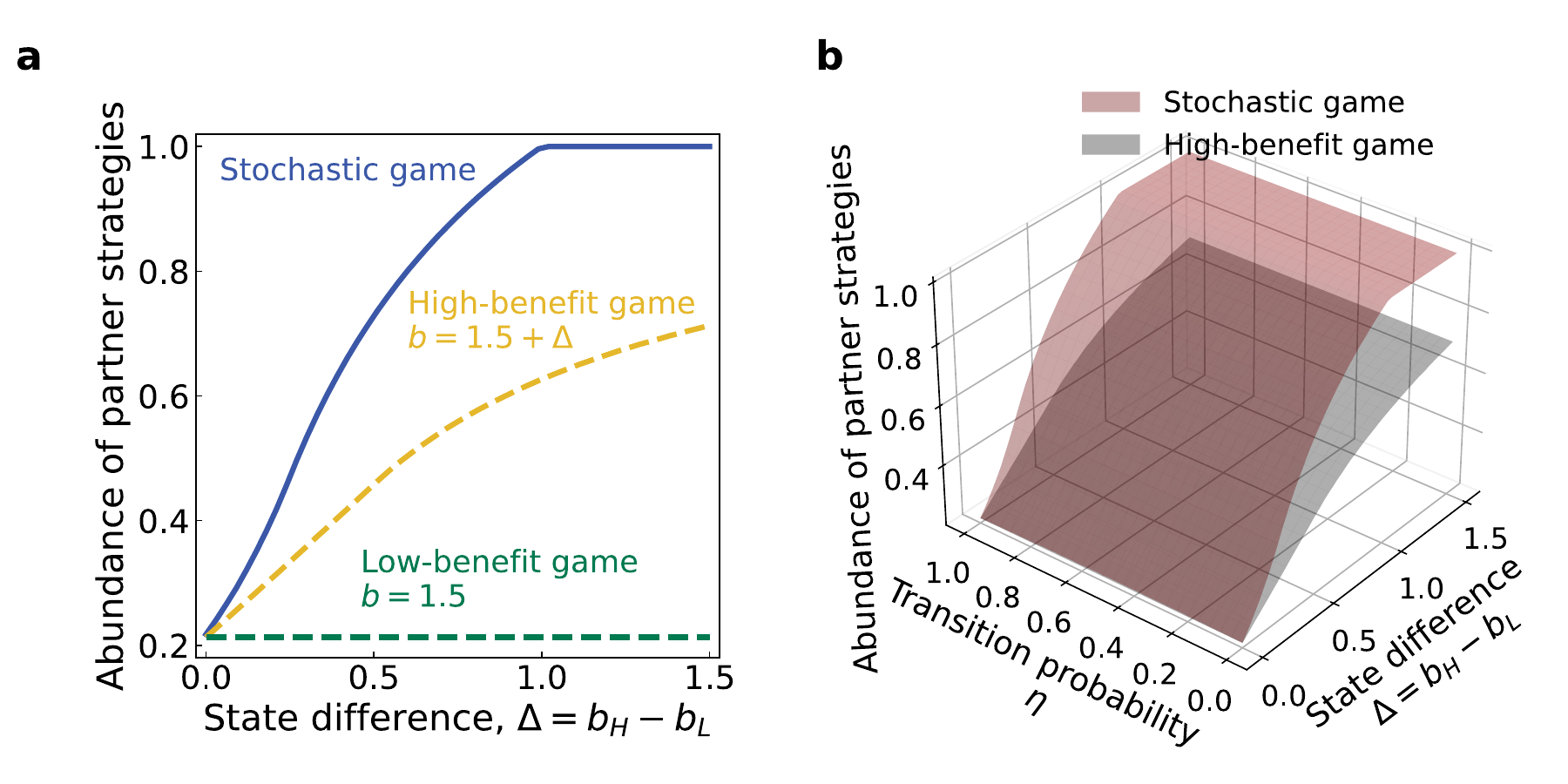}
	\caption{\textbf{Environmental feedback promotes the stability of cooperative strategies.} \textbf{a}, Abundance of partner strategies as a function of the state difference $\Delta$. Dashed lines show abundance in asynchronous static games as baselines, compared to asynchronous stochastic games (solid line). When $\Delta \ge 1$, the abundance reaches $100\%$.
    \textbf{b}, Variation in the abundance of partner strategies with state difference $\Delta$ and transition probability $\eta$ in asynchronous stochastic (red plane) and asynchronous high-benefit (gray plane) games. Results for asynchronous low-benefit games are omitted but can be derived by setting $\Delta = 0$ in the gray plane. Across all values of $\Delta$ and $\eta$, asynchronous stochastic games consistently show higher abundance of partner strategies. Parameter values: $b_L=1.5$, $c=1$, $\bm{\lambda}=(1,0,\eta,0)$, and $\eta=0.5$ for \textbf{a}.}
	\label{fig_3}
\end{figure}

In asynchronous static games, partner strategies never completely dominate the cooperative strategy space. This occurs because, for any given benefit $b$ and cost $c$, there always exist cooperative strategies that fail to satisfy the partner condition specified by Eq.~(\ref{partner_condition_static}). For instance, when $p_{CD}$ and $p_{DC}$ fall within the range $(b/(b+c), 1)$, the cooperative strategy $\mathbf{p}=(1,p_{CD},p_{DC},p_{DD})$ does not meet the last inequality of the partner condition. As a result, the abundance remains below $1 - c^2/(b+c)^2$. 
In contrast, asynchronous stochastic games allow partner strategies to fully dominate the cooperative strategy space. When the benefit difference between high and low states ($\Delta = b_H - b_L$) exceeds the cooperation cost $c$, any cooperative strategy qualifies as a partner strategy. This occurs due to environmental feedback, which amplifies the benefits of long-term cooperation and reinforces sustained cooperative behavior.
Thus, dynamic environments encourage a higher prevalence of partner strategies than static ones, as illustrated in Fig.~\ref{fig_3}. Moreover, as the benefit difference between states increases, the abundance of partner strategies grows, indicating that larger benefit disparities promote the stability of reciprocal behavior in asynchronous stochastic games.

\subsection*{Emergence of cooperation in evolutionary asynchronous stochastic games}

In the preceding analysis, we examined the interaction between two specific individuals and explored the conditions that enable them to maintain stable reciprocity. We now shift focus to explore how asynchronous interaction affects the emergence of cooperation in an evolving population. Our model of evolutionary asynchronous stochastic games involves a pairwise comparison process within $N$ individuals. Each participant employs a memory-1 strategy and accumulates payoffs through interactions with all other members of the population. Strategy adjustment occurs through two mechanisms: strategy imitation and random exploration. In strategy imitation, players are more likely to adopt strategies that yield higher payoffs, a tendency governed by the selection strength $\beta > 0$. Concurrently, random exploration introduces mutation, enabling players to choose new strategies from the full set of memory-1 strategies. For more details, refer to the Materials and Methods section. For comparison, we also include the results from evolutionary asynchronous static games.

\begin{figure}[!t]
	\centering
	\includegraphics[width=.8\textwidth]{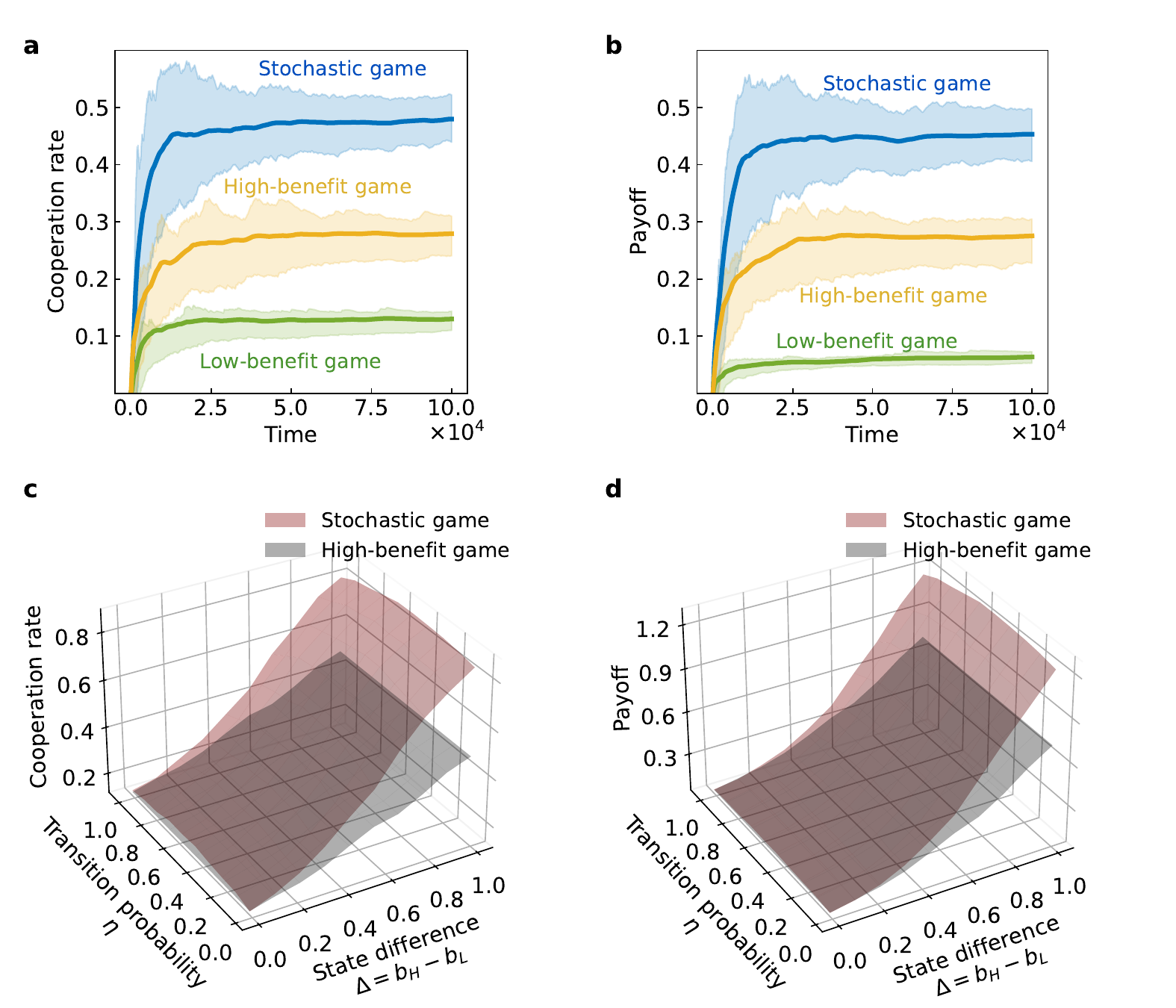}
	\caption{\textbf{Emergence of cooperation in evolutionary asynchronous stochastic games.} \textbf{a}, \textbf{b}, Changes in global cooperation rate and payoff over time for the asynchronous stochastic (blue), asynchronous high-benefit (yellow), and asynchronous low-benefit (green) games. The time axis corresponds to the number of mutant strategies produced within the population (see Materials and Methods). The solid lines depict the average outcomes from 100 simulations, with shaded areas indicating the interquartile range. The population in the asynchronous stochastic game consistently achieves the highest cooperation rates and payoffs. \textbf{c}, \textbf{d}, Joint effects of state difference $\Delta$ and transition probability $\eta$ on evolution outcomes in the asynchronous stochastic (red plane) and asynchronous high-benefit (gray plane) games. These plots are obtained by interpolating the original data. Across all values of $\Delta$ and $\eta$, environmental feedback enhances cooperation rates and payoffs for the population. Parameter values: $b_H = 2$ (\textbf{a} and \textbf{b}), $b_L = 1.5$, $c = 1$, $\bm{\lambda} = (1,0,\eta,0)$, $\eta = 0.5$ (\textbf{a} and \textbf{b}), $\beta = 1$, $\varepsilon=0$, $\mu \rightarrow 0$, and $N = 50$.}
	\label{fig_5}
\end{figure}

Figures~\ref{fig_5}\textbf{a} and~\textbf{b} show the global cooperation rates and payoffs throughout the evolutionary process. The cooperation rate in the asynchronous stochastic game ($48\%$) surpasses that in static contexts ($28\%$ and $13\%$), leading to a higher payoff of $0.45$, compared to $0.28$ and $0.06$ in asynchronous static games. These findings highlight the critical role of feedback mechanisms in sustaining cooperation and enhancing social welfare within asynchronous interactions.
To assess the robustness of these findings, we examine how variations in key parameters affect interaction outcomes. Figures~\ref{fig_5}\textbf{c} and~\textbf{d}, along with Fig. S3, demonstrate that evolutionary feedback consistently promotes cooperative behavior. In scenarios involving large populations, strong selection, and low mutation, these feedback loops also improve overall welfare. The advantages of evolutionary feedback remain robust even in the presence of implementation errors (Fig. S4) and are further supported by a theoretical approximation under weak selection ($\beta \rightarrow 0$) (Fig. S5).

In evolutionary asynchronous stochastic games, cooperation emerges due to a dual ``shadow of the future'': (1) individuals who choose to defect risk facing retaliatory actions in subsequent rounds; (2) engaging in defective behavior can harm the environment, potentially reducing future payoffs. The second shadow operates differently in asynchronous versus synchronous interactions. In synchronous games, both players share the same environmental conditions each round, so any environmental harm affects both equally. In contrast, in asynchronous games, environmental states can vary between players within the same round, making the consequences of defection more uneven. Specifically, the timing of interactions may allow defectors to incur fewer negative repercussions than their counterparts. This imbalance adds complexity to strategic decision-making, as players might exploit the sequence to gain immediate benefits while shifting a greater share of environmental costs onto others. Consequently, cooperative behavior in asynchronous interactions within dynamic environments may be more vulnerable—a complexity we explore further in the next subsection.

\subsection*{Cooperation-inhibiting effects of interaction asynchrony}

\begin{figure}[!t]
	\centering
	\includegraphics[width=.6\textwidth]{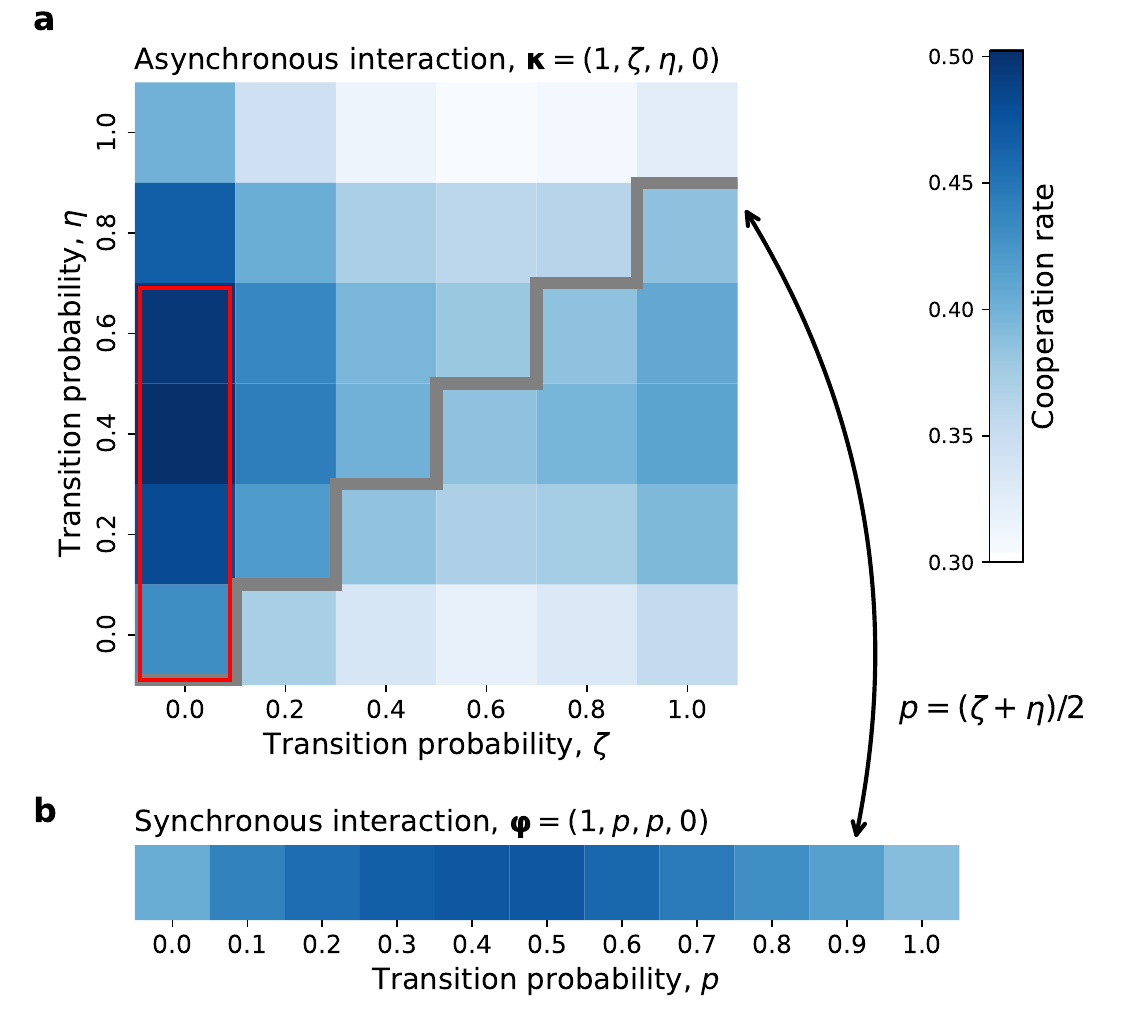}
	\caption{\textbf{Interaction asynchrony inhibits the evolution of cooperation.} 
    Comparison of cooperation rates in the asynchronous and synchronous stochastic games. \textbf{a}, In the asynchronous stochastic game, the transition vector is $\bm{\kappa}=(1,\zeta,\eta,0)$. The region above the gray line satisfies the constraint $\eta \ge \zeta$, which demonstrates the reality that the impact of players' actions diminishes over time. \textbf{b}, In the synchronous stochastic game, the transition vector is $\bm{\varphi}=(1,p,p,0)$. 
    We set $p=(\zeta+\eta)/2$ to compare the cooperation rates in synchronous interactions with those in asynchronous interactions. 
    Overall, asynchronous interactions tend to suppress cooperation relative to synchronous interactions, except when $\zeta=0$ and $\eta$ is moderate or small (the region highlighted by the red box).
    Notably, the red box lies above the gray line. 
    Default parameters: $b_H = 2$, $b_L = 1.5$, $c = 1$, $\beta = 1$, $\varepsilon = 0$, $\mu \rightarrow 0$, and $N = 50$.}
	\label{fig_8}
\end{figure}

\begin{figure}[!t]
	\centering
	\includegraphics[width=.8\textwidth]{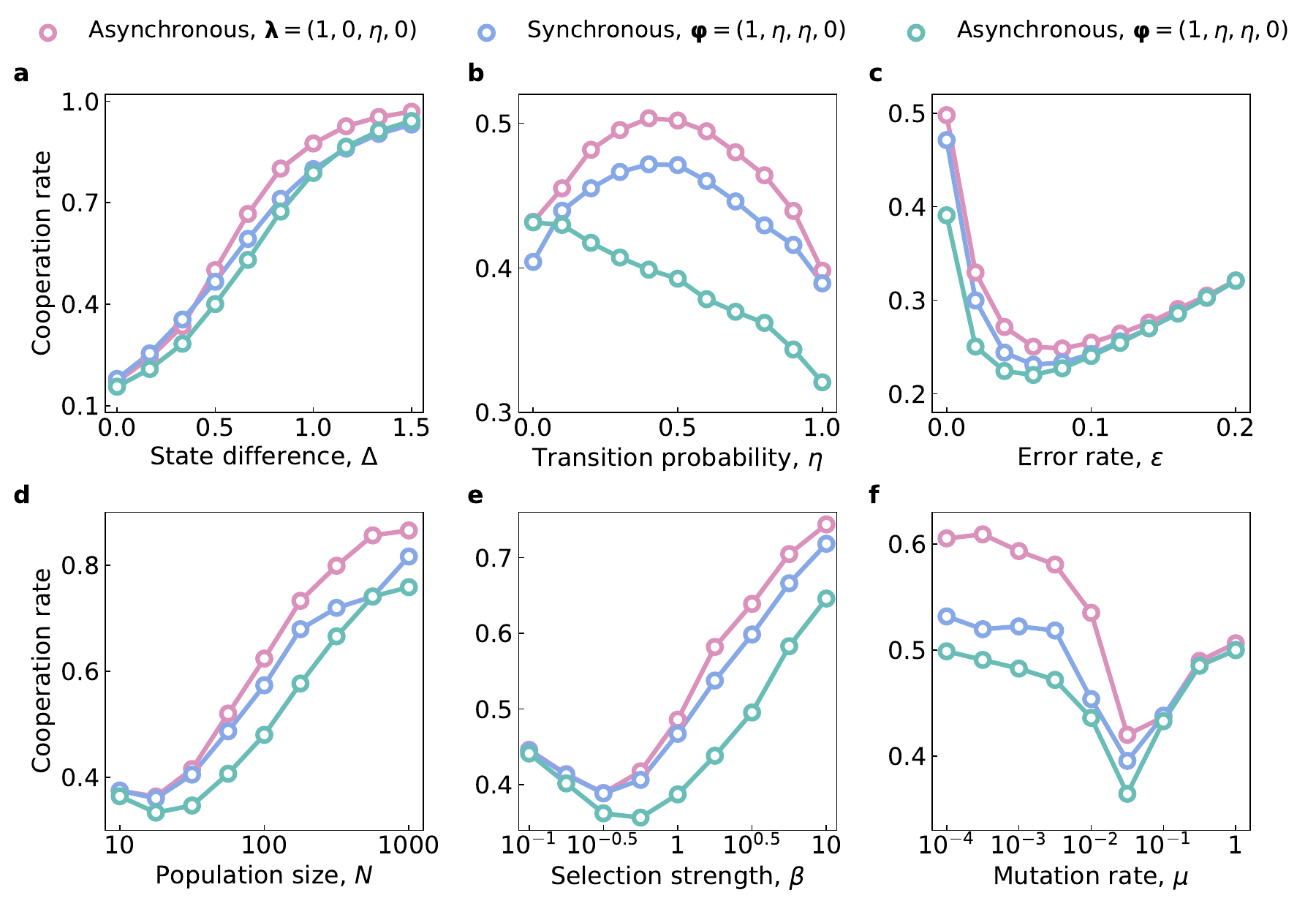}
	\caption{\textbf{The inhibitory effects of asynchronous interactions are robust with respect to parameter changes.} 
    We study asynchronous stochastic games with transition vector $\bm{\lambda} = (1, 0, \eta, 0)$, asynchronous stochastic games with transition vector $\bm{\varphi} = (1, \eta, \eta, 0)$, and synchronous stochastic games with transition vector $\bm{\varphi} = (1, \eta, \eta, 0)$.
    The cooperation rates are presented as a function of state difference $\Delta$ (\textbf{a}), transition probability $\eta$ (\textbf{b}), error rate $\varepsilon$ (\textbf{c}), population size $N$ (\textbf{d}), selection strength $\beta$ (\textbf{e}), and mutation rate $\mu$ (\textbf{f}). Cooperation is enhanced by large state differences, larger population sizes, and stronger selection, but reduced by high error and mutation rates. Moderate transition probabilities tend to support cooperation, with the exception of asynchronous games with $\bm{\varphi}$, where higher $\eta$ reduces cooperation. When both interaction modes use $\bm{\varphi}$, the synchronous one leads to higher cooperation rates. Default parameters: $b_H = 2$, $b_L = 1.5$, $c = 1$, $\eta=0.5$, $\beta = 1$, $\varepsilon = 0$, $\mu \rightarrow 0$, and $N = 50$.}
	\label{fig_7}
\end{figure}

Previous studies have shown that asynchronous and synchronous interactions exhibit similar cooperative dynamics \cite{nowak1994alternating,park2022cooperation}. However, these studies overlook that real-world environments are typically dynamic, both influencing and being influenced by interactions themselves. Therefore, it is crucial to investigate whether asynchronous interactions lead to distinct patterns of cooperative dynamics compared to synchronous interactions in such dynamic environments.
In synchronous stochastic games, we adopt the transition vector $\bm{\varphi} = (1, p, p, 0)$, where $p$ denotes the probability of transitioning to a high-benefit state when one player cooperates and the other defects, aligning with established conventions in studies of synchronous interactions \cite{hilbe2018evolution}. In contrast, asynchronous stochastic games utilize the transition vector $\bm{\kappa} = (1, \zeta, \eta, 0)$, capturing the temporal dynamics of asynchronous interactions. 

Figure~\ref{fig_8} presents a comparative analysis of cooperation rates between asynchronous and synchronous stochastic games. For fair comparison, each $(\zeta, \eta)$ pair in asynchronous games is matched with a transition probability $p = (\zeta + \eta)/2$ in synchronous games. This approach ensures that after one cooperation and one defection, the average probability of transitioning to a high-benefit state is the same for both interaction types. 
The results reveal that the order of interaction plays a critical role in determining cooperation rates: asynchronous interactions generally impede cooperation, except under specific conditions where $\zeta = 0$ and $\eta$ is moderate or low, as highlighted in the red box of Fig.~\ref{fig_8}. This exception occurs within the region where $\eta \geq \zeta$, mirroring real-world scenarios where cooperative behavior more effectively preserves environmental resources than defection, and the influence of a player's behavior diminishes over time. 
Notably, the transition vector $\bm{\lambda} = (1, 0, \eta, 0)$, which is employed in earlier subsections, precisely aligns with the conditions outlined by the red box, underscoring its relevance in fostering cooperation within asynchronous interactions.

To further evaluate the robustness of the cooperation-inhibiting effects of asynchronous interactions, we vary key parameters. By redefining $\bm{\varphi} = (1, \eta, \eta, 0)$, we examine cooperation rates in asynchronous stochastic games with both $\bm{\lambda}$ and $\bm{\varphi}$ transition vectors, as well as in synchronous stochastic games using $\bm{\varphi}$. Figure~\ref{fig_7} illustrates these cooperation rates, revealing that cooperation thrives with larger state differences, increased population sizes, and stronger selection strength, but it suffers from higher error and mutation rates. Additionally, in asynchronous interactions, cooperation rates are higher when transitions follow $\bm{\lambda} = (1, 0, \eta, 0)$ compared to $\bm{\varphi} = (1, \eta, \eta, 0)$. This disparity arises because $\bm{\lambda}$ enforces a stricter feedback loop by ensuring $\lambda_{DC} > \lambda_{CD}$, thereby emphasizing the more detrimental impact of recent defections on environmental resources. Such asymmetry aligns better with the nature of asynchronous interactions, thereby promoting higher levels of cooperation. When transitions follow $\bm{\varphi}$ in both interaction types, cooperation is more common in synchronous games than in asynchronous settings.

The reduction in cooperation rates within asynchronous interactions can be primarily attributed to their asynchrony. Synchronous decision-making processes mitigate uncertainty and build trust, creating an equitable environment conducive to cooperative behavior. In contrast, the staggered decision-making characteristic of asynchronous interactions introduces misalignment and unpredictability, making it challenging for participants to accurately anticipate each other's actions. This uncertainty breeds mistrust and hesitation, impeding cooperative behavior. 
To overcome the challenges posed by asynchrony in asynchronous interactions, robust feedback mechanisms are necessary. Transition vectors like $\bm{\lambda} = (1, 0, \eta, 0)$, which conform to the nature of asynchronous interactions, effectively maintain high cooperation levels. By reinforcing the negative consequences of defection and aligning incentives with the order of interactions, $\bm{\lambda}$ ensures that recent defections have a more pronounced detrimental effect on environmental resources than earlier ones. These tailored mechanisms boost cooperation despite the inhibitory effects of asynchronous interactions.

\section{Discussion}
Cooperation is crucial in both human and animal societies, and understanding its persistence across various contexts has been a long-standing research focus. Direct reciprocity, where individuals make decisions based on previous interactions, is a key factor that supports cooperative behavior. While most studies focus on synchronous interactions, many real-world scenarios involve turn-taking, such as vampire bats sharing blood \cite{wilkinson1984reciprocal} and pied babblers acting as sentinels \cite{ridley2013sentinel}. This asynchrony reshapes decision-making processes, as strategies effective in synchronous settings may falter in asynchronous contexts \cite{nowak1994alternating, frean1994prisoner}. Additionally, environments are inherently dynamic; individual behaviors continuously shape the environment, which in turn influences future interactions. Given the prevalence and impact of asynchronous interactions, it is important to study direct reciprocity in this framework. To address this gap, we introduce an asynchronous stochastic game model where participants alternate in making decisions, and these decisions influence the environment. By emphasizing asynchronous interactions in a dynamic setting, our framework more accurately captures the complexities of real-world interactions.

Our analysis identifies partner strategies in asynchronous stochastic games. These strategies ensure full cooperation in self-play and constitute Nash equilibria, promoting ongoing cooperative behavior. 
We examine the prevalence of partner strategies in the cooperative strategy space, as a higher prevalence indicates greater stability of cooperative behavior in asynchronous interactions. Our analysis shows that partner strategies are more prevalent in asynchronous stochastic games than in static ones. This increased prevalence is significantly influenced by the difference in benefits between states, which plays a critical role in maintaining cooperation.
When the variation in benefits between environmental states exceeds the cost of cooperation, stable reciprocal relationships become ubiquitous. In such scenarios, if both players continue to cooperate after mutual cooperation, neither has an incentive to deviate, thereby ensuring the sustainability of reciprocity. This universal reciprocity is unattainable in asynchronous static games, where cooperative strategies may lack stability.

We explore how asynchronous interactions impact the emergence of cooperation by studying the evolutionary dynamics of populations engaged in asynchronous games. We present results for both evolutionary asynchronous stochastic and static games.
Our results indicate that dynamic environments, which are influenced by players' actions and, in turn, influence those actions, encourage higher levels of cooperative behavior in asynchronous interactions compared to static environments. In dynamic environments, defection not only leads to future punishments from co-players but also harms the environment, resulting in reduced future payoffs. This degradation deters short-term selfish behavior, fostering a culture of cooperation to protect environmental integrity and ensure sustained long-term benefits. 
Additionally, in large populations with strong selection strength and low mutation rates, asynchronous stochastic games yield higher payoffs than their static counterparts.

We investigate how cooperative behavior differs between evolutionary asynchronous and synchronous stochastic games. Our findings reveal that the order of interactions plays a crucial role in shaping cooperative dynamics, with asynchronous interactions generally hindering cooperative behavior. This contrasts with earlier research suggesting similar patterns of cooperation in both asynchronous and synchronous interaction modes \cite{nowak1994alternating, park2022cooperation}.
The key difference lies in our consideration of a dynamic environment, as opposed to the static environments addressed in previous studies. In synchronous games, players act synchronously in a shared environment, allowing them to align strategies and develop mutual expectations within a symmetric game framework. However, in asynchronous interactions, asynchrony creates a time gap between actions, disrupting the flow of information. Furthermore, the environment can differ for players even within the same round, resulting in asymmetric games that are more variable and less predictable. This interplay of asymmetry and asynchrony increases uncertainty, making it difficult for players to anticipate others' actions and adjust their strategies effectively. Consequently, cooperation is typically more challenging to evolve in asynchronous interactions compared to synchronous ones.
However, we also identify significant exceptions. Specifically, cooperation can flourish under asynchronous interactions when tailored transition vectors are employed. These feedback mechanisms effectively mitigate the negative impacts of asynchrony by reinforcing the consequences of defection in a manner that aligns with the temporal nature of asynchronous decision-making. Our results underscore the potential of designing strategic transition vectors to promote cooperation despite the challenges posed by asynchrony and environmental variability.

This study offers two main contributions. First, we identify partner strategies in asynchronous stochastic games, demonstrating that environmental feedback promotes the stability of cooperative strategies, supporting cooperation in asynchronous interactions. Second, we observe that interaction asynchrony hinders cooperation, underscoring the challenges in fostering cooperation in asynchronous contexts. Our research paves the way for future studies. While we focus on state-independent transitions—where the subsequent environmental state does not depend on the current one—future work could explore state-dependent transitions to deepen our understanding of cooperation in dynamic environments. Additionally, although our model assumes strictly asynchronous games, real-world interactions are often irregular, such as when a vampire bat obtains blood on consecutive nights while another fails, leading to unilateral reciprocity. Incorporating irregular rotations could better simulate real-world interactions and provide insights into differences between asynchronous and synchronous interactions. These extensions will advance theory and enhance our understanding of the evolution of cooperation in biological and social systems.

\section{Methods}
\subsection*{Calculation of cooperation rate and payoff}
We begin by considering the error-free asynchronous stochastic game ($\varepsilon=0$). Denote the strategies of Players 1 and 2 as $\mathbf{p}$ and $\mathbf{q}$, respectively. Here, $p_{a\tilde{a}}^s$ (resp. $q_{a\tilde{a}}^s$) denotes the probability for Player 1 (resp. Player 2) to cooperate given the current environmental state $s$, the player's own previous action $a$ two periods ago, and the co-player's action $\tilde{a}$ in the last period. The interaction between these strategies can be represented by a Markov chain. The state space is $\{s_1, s_2, a_1, a_2\} \subseteq \{H, L\}^2 \times \{C, D\}^2$, where $a_i$ is Player $i$'s action, and $s_i$ is the environmental state.
The frequency of the system being in these states is calculated by solving for the unique stationary distribution $\mathbf{v}$, which satisfies $\mathbf{v} = \mathbf{v} \mathbf{M}$. Here, $\mathbf{M}$ is the transition matrix, and  $\mathbf{v}$ exists if the Markov chain is irreducible and aperiodic. If these conditions are not met, the long-term outcomes of the game depend on the initial actions of both players. The cooperation rate $\mathcal{C}(\mathbf{p},\mathbf{q})$ for Player 1 is defined as the weighted average of the cooperation rates in these states, with the weights given by $\mathbf{v}$. The corresponding payoff $\pi(\mathbf{p},\mathbf{q})$ is defined analogously.

These calculations can be extended to error-prone scenarios, where a player chooses the opposite action with probability $\varepsilon>0$. In this context, Players $1$ and $2$ employ modified memory-1 strategies, $\mathbf{p}^\varepsilon = \varepsilon + (1-2\varepsilon)\mathbf{p}$ and $\mathbf{q}^\varepsilon = \varepsilon + (1-2\varepsilon)\mathbf{q}$, which ensure that the Markov chain is irreducible and aperiodic, thus guaranteeing a unique stationary distribution.

\subsection*{Evolutionary asynchronous stochastic games}
We model evolutionary dynamics in a well-mixed population of $N$ individuals, where each player adopts a memory-1 strategy for the asynchronous stochastic game. A player's payoff is the average result from interactions with all other members in the population. At each time step, one player is randomly chosen to update strategy. With probability $\mu$, the player performs random exploration, randomly choosing a new strategy from the entire memory-1 strategy set. Otherwise, the player randomly selects a role model player from the population, and imitates the role model player's strategy with probability given by:
\begin{equation}
	\frac{1}{1+\exp \left[-\beta \left(\pi_{R}-\pi_{F}\right)\right]},
\end{equation}
where $\pi_F$ (resp. $\pi_R$) denotes the payoff of the focal player (resp. the role model player), and $\beta$ is the selection strength.

In most cases, we assume the mutation rate $\mu \rightarrow 0$, implying that a mutant strategy either becomes fixed or goes extinct before another mutation arises. This assumption enhances simulation efficiency and enables theoretical approximation of the cooperation rate.

\subsection*{Theoretical approximation of the cooperation rate}
Our theoretical approximation operates under the assumption of rare mutation ($\mu \rightarrow 0$) and weak selection ($\beta \rightarrow 0$). We sample $K$ strategies from the complete set of memory-1 strategies, referring to this subset as $\Omega$. The fixation probability of strategy $A$ among $N-1$ individuals using strategy $B$ can be calculated as $\rho_{AB} = 1/N + \beta \gamma_{AB} + O(\beta^2)$. With rare mutations, the dynamics of the population can be characterized by a transition matrix $\mathbf{P}$ defined on the state space $\Omega$. The stationary distribution $\mathbf{u}$ of the transition matrix $\mathbf{P}$ describes the time the population spends in each state. For weak selection, the element $u_{\mathbf{p}}$, corresponding to state $\mathbf{p}$ within vector $\mathbf{u}$, can be expressed as \cite{tarnita2011multiple}:
\begin{equation}\label{approx_v}
	u_{\mathbf{p}}=\frac{1}{K} + \beta \frac{N}{K^{2}} \sum_{\mathbf{q} \in \Omega} (\gamma_{\mathbf{p}\mathbf{q}} - \gamma_{\mathbf{q}\mathbf{p}}) + O(\beta^2).
\end{equation}
Under rare mutation, the cooperation rate is given by
\begin{equation}
	\langle \mathcal{C} \rangle = \sum_{\mathbf{p} \in \Omega} \mathcal{C}_{\mathbf{p}} \cdot u_{\mathbf{p}},
\end{equation}
where $\mathcal{C}_{\mathbf{p}}=\mathcal{C}(\mathbf{p},\mathbf{p})$ specifies the cooperation rate when the strategy $\mathbf{p}$ plays against itself. For weak selection, we have
\begin{equation}
	\begin{aligned}
		\langle \mathcal{C} \rangle =& \sum_{\mathbf{p} \in \Omega} \mathcal{C}_{\mathbf{p}} \left[ \frac{1}{K} + \beta \frac{N}{K^{2}} \sum_{\mathbf{q} \in \Omega} (\gamma_{\mathbf{p}\mathbf{q}} - \gamma_{\mathbf{q}\mathbf{p}})  + O(\beta^2)\right]\\
		=& \frac{1}{K}\sum_{\mathbf{p} \in \Omega} \mathcal{C}_{\mathbf{p}}+\frac{\beta N}{K^2}\sum_{\mathbf{p} \in \Omega}\sum_{\mathbf{q} \in \Omega} \mathcal{C}_{\mathbf{p}} (\gamma_{\mathbf{p}\mathbf{q}} - \gamma_{\mathbf{q}\mathbf{p}}) + O(\beta^2).
	\end{aligned}
\end{equation}
Since the expected value of the first term is $1/2$, representing the cooperation rate under neutral drift, we focus on the coefficient of the first-order term of $\beta$.

\bibliographystyle{unsrt}
\bibliography{reference}

\end{document}